\documentstyle[11pt]{article}
\renewcommand{\theequation}{\thesection.\arabic{equation}}
\renewcommand{\thefootnote}{\fnsymbol{footnote}}
\newlength{\extraspace}
\newlength{\extraspaces}
\newcommand{\be}{\begin{equation}
\addtolength{\abovedisplayskip}{\extraspaces}
\addtolength{\belowdisplayskip}{\extraspaces}
\addtolength{\abovedisplayshortskip}{\extraspace}
\addtolength{\belowdisplayshortskip}{\extraspace}}
\newcommand{\ee}{\end{equation}}

\newcommand{\ba}{\begin{eqnarray}
\addtolength{\abovedisplayskip}{\extraspaces}
\addtolength{\belowdisplayskip}{\extraspaces}
\addtolength{\abovedisplayshortskip}{\extraspace}
\addtolength{\belowdisplayshortskip}{\extraspace}}
\newcommand{\ea}{\end{eqnarray}}

\newcommand{\bas}{\begin{eqnarray*}
\addtolength{\abovedisplayskip}{\extraspaces}
\addtolength{\belowdisplayskip}{\extraspaces}
\addtolength{\abovedisplayshortskip}{\extraspace}
\addtolength{\belowdisplayshortskip}{\extraspace}}
\newcommand{\eas}{\end{eqnarray*}}

\newcounter{subequation}[equation]
\makeatletter

\expandafter\let\expandafter\reset@font\csname reset@font\endcsname
\def\subeqnarray{\arraycolsep1pt
    \def\@eqnnum\stepcounter##1{\stepcounter{subequation}
        {\reset@font\rm(\theequation\alph{subequation})}}\eqnarray}

\makeatother

\def\eql#1{\\(#1)\vadjust{\penalty10000\vskip-5.3ex}}

\newcommand{\newsection}[1]{
\vspace{10mm}
\pagebreak[3]
\addtocounter{section}{1}
\setcounter{equation}{0}
\setcounter{subsection}{0}
\setcounter{footnote}{0}

\begin{flushleft}
{\large\bf \thesection. #1}
\end{flushleft}
\nopagebreak
\medskip
\nopagebreak}

\newcounter{prop}

\newcommand{\NP}[1]{Nucl.\ Phys.\ {\bf #1}}

\newcommand{\CMP}[1]{Comm.\ Math.\ Phys.\ {\bf #1}}

\newcommand{\N}{\mbox{I\hspace{-.4ex}N}}
\newcommand{\C}{\mbox{$\,${\sf I}\hspace{-1.2ex}{\bf C}}}
\newcommand{\Z}{\mbox{{\sf Z}\hspace{-1ex}{\sf Z}}}
\newcommand{\R}{\mbox{\rm I\hspace{-.4ex}R}}
\newcommand{\1}{\mbox{1\hspace{-.6ex}1}}
\newcommand{\bra}{\langle}
\newcommand{\ket}{\rangle}
\newcommand{\ra}{\rightarrow}

\newcommand{\is}{\! & \! = \! & \!}
\newcommand{\nonum}{\nonumber \\[1.5mm]}
\newcommand{\sspace}{\makebox[1cm]{ }}
\newcommand{\bspace}{\makebox[2cm]{ }}

\newcommand{\sh}{{\rm sh\,}}
\newcommand{\ch}{{\rm ch\,}}
\renewcommand{\tanh}{\rm th\,}
\newcommand{\th}{\theta}

\newcommand{\dd}[1]{ \frac{\partial}{\partial{#1}} }

\newcommand{\cO}{{\cal O}}

\newcommand{\Fb}{{F^{(\beta)}_{ab}}}

\begin{document}
%
\begin{titlepage}
%
\renewcommand{\thefootnote}{\fnsymbol{footnote}}
\begin{flushright}
MPI-PhT/94-73\\
hep-th/9503139
\end{flushright}
\vspace{1cm}

\begin{center}
{\LARGE \makebox[3mm]{ }
On the Free Field Realization of Form Factors}
\vspace{2cm}

{\large Max R. Niedermaier}\\ [3mm]
{\small\sl Max-Planck-Institut f\"{u}r Physik}\\
{-\small\sl Werner Heisenberg Institut - }\\
{\small\sl F\"{o}hringer Ring 6}\\
{\small\sl D-80805 Munich, Germany}
\vspace{3.5cm}

{\bf Abstract}
\end{center}
\begin{quote}
A method to construct free field realizations for the
form factors of diagonal factorized scattering theories
is described. Form factors are constructed from linear functionals
over an associative `form factor algebra', which in particular
generate solutions of the deformed Knizhnik-Zamolodchikov equations
with parameter $2\pi$. We show that there exists a unique deformation
off the (`Rindler') value $2\pi$ that preserves the original
$S$-matrix and which allows one to realize form factors
as vector functionals over an algebra of generalized vertex
operators.
\end{quote}
\vfill
\renewcommand{\thefootnote}{\arabic{footnote}}
\setcounter{footnote}{0}
\end{titlepage}

\newsection{Introduction}
The form factor bootstrap \cite{KarWeisz,Smir1} is a universal
approach to integrable relativistic QFTs. It takes an
implementation of the Wightman axioms in terms of form factors as a
starting point. As a consequence of the factorized scattering theory
there exists a recursive system of coupled Riemann-Hilbert equations
for these form factors, which entail that the Wightman functions
built from them have all the desired properties. However, the development
of systematic solution techniques for these equations is still in its
beginnings. An algebraic solution scheme was proposed in \cite{MN1}:
There exists a doublet of form factor algebras $F_{\pm}(S)$ associated
with a given two particle $S$-matrix such that suitable (`$T$-invariant')
linear functionals over these algebras define `pre-'form factors
$f^{\pm}$. The sum $f=f^+ + f^-$ then automatically
satisfies all the form factor axioms, except for the bound state
residue axiom. The construction of form factors thus reduces to
constructing a realization of $F_{\pm}(S)$ and to find $T$-invariant
functionals in this realization. Applied to the case of integrable
QFTs with a diagonal factorized scattering theory, the construction
also lead to a universal formula for the eigenvalues of the local
conserved charges. The purpose of the present paper is to show
how this set-up relates to the vertex operator construction of
the eigenvalues \cite{MN2} and to the proposal
\cite{Kyoto,Luk} to realize form factors as (trace) functionals over
an algebra of generalized vertex operators. The functionals in question
are ill defined in general, so that some kind of regularization
is required. We propose to construct form factors as limits
\be
f_{a_n\ldots a_1}(\th_n,\ldots, \th_1) =\lim_{\beta\ra 2\pi}\,
f^{(\beta)}_{a_n\ldots a_1}(\th_n,\ldots, \th_1)\;,
\ee
where the functions $f^{(\beta)}_{a_n\ldots a_1}(\th_n,\ldots, \th_1)$
solve a system of deformed form factor equations. The deformation is
uniquely determined by the following two conditons:
\begin{itemize}
\item[$i.$] The `deformed form factors'
$f^{(\beta)}_{a_n\ldots a_1}(\th_n,\ldots, \th_1)$ solve the
deformed Knizhnik-Za\-mo\-lodchi\-kov equations (KZE) with
parameter $i\beta$, i.e.
\ba
f^{(\beta)}_{a_n\ldots a_1}(\th_n,\ldots,\th_k+i\beta,\ldots,\th_1) \is
(A^{(\beta)}_k)_{a_n\ldots a_1}(\th_n,\ldots,\th_1)\,
f^{(\beta)}_{a_n\ldots a_1}(\th_n,\ldots, \th_1)\;,\nonum
(A^{(\beta)}_k)_{a_n\ldots a_1}(\th_n,\ldots,\th_1) \is
\prod_{j<k}S_{a_ja_k}(\th_{jk})
\prod_{j>k}S_{a_ka_j}(\th_{jk}-i\beta)\;,
\ea
for the case of a diagonal $S$-matrix, considered here.
\item[$ii.$] The factorized scattering theory (i.e.~the two particle
S-matrix) remains undeformed. In particular, the deformed form factors
satisfy for  $Re\,\th_{k+1,k}\neq 0$
\ba
&& f^{(\beta)}_{a_n\ldots a_{k+1}a_k\ldots  a_1}
(\th_n,\ldots,\th_{k+1},\th_k,\ldots,\th_1) \nonum
&&\sspace = S_{a_ka_{k+1}}(\th_{k+1,k})\,
f^{(\beta)}_{a_n\ldots a_ka_{k+1}\ldots  a_1}
(\th_n,\ldots,\th_k,\th_{k+1},\ldots,\th_1)\;.
\ea
\end{itemize}
We shall refer to this deformation as `S-matrix preserving deformation'.
The original value $\beta=2\pi$ has special features in many respects.
In particular, it indirectly reflects the thermal properties of the
Rindler vacuum. For $\beta \neq 2\pi$ the functions
$f^{(\beta)}_{a_n\ldots a_1}(\th_n,\ldots, \th_1)$
characterize a `deformed' relativistic QFT. We shall make some
suggestions on the physical significance of this deformation at the end
of the paper. Here we shall be concerned with a purely technical
consequence of taking $\beta \neq 2\pi$: The deformation serves as
a regularization, in the sense that for
$\beta/2\pi$ irrational the deformed `pre-'form factors $f^{(\beta,\pm)}$
can be realized as vector functionals over a realization of a deformed
form factor doublet $F^{(\beta)}_{\pm}(S)$. The realization is in terms
of generalized vertex operators
\be
f^{(\beta,\pm)}(X^{\pm}) =\bra\cO|\rho^{(\beta)}(X^{\pm})|0\ket\;,
\sspace X^{\pm}\in F^{(\beta)}_{\pm}(S)\;,
\ee
where $|0\ket$ is the vacuum of a tensor product of free bosonic Fock
spaces and $|\cO\ket$ is a state characterizing the local operator
in question. The functionals (1.4) could also be written as trace
functionals, in which case the above deformation also serves to
regularize the trace. Compared to other regularizations \cite{Luk} it
has the advantage to preserve the conserved charge bootstrap i.e. the
relation between the position of the bound state poles in the S-matrix
and the spins of the local conserved charges (c.f. section 3).

\newsection{S-matrix preserving deformation of form factors}
Suppose a factorized scattering theory to be given with
a diagonal $S$-matrix satisfying the bootstrap equations
\ba
&& S_{ab}(\th)=S_{ba}(\th) =S_{ab}(-\th)^{-1}=
S_{ab}^{*}(-\th^*)\nonum
&& S_{ab}(i\pi -\th) = S_{a\bar{b}}(\th)\nonum
&& S_{da}(\th +i\eta(a))\,S_{db}(\th + i\eta(b))\,
S_{dc}(\th +i\eta(c)) =1 \;.
\ea
Here $S_{ab}(\th)$ is the $S$-matrix element for the
elastic scattering of particles of type $a,b \in
\{1,\ldots ,r\}$. It is a periodic meromorphic function in the
rapidity difference $\th =\th_a-\th_b$ with
period $2\pi i$. Bound state poles are situated on the
physical sheet $0 \leq Im\,\th <\pi$.
The first equation expresses hermitian analyticity and
(formal) unitarity. The second equation implements crossing
invariance in terms of a charge conjugation operation
$a\ra \bar{a}\in \{1,\ldots ,r\}$. The last equation is the
the bootstrap equation proper, where the triplet
$(\eta(a),\eta(b),\eta(c))$ is related to the conventional
fusing angles. It is convenient to introduce the scattering phase
adopted to the normalization condition at $\th =0$
\be
S_{ab}(\th)=\epsilon_{ab}\,e^{\textstyle i\delta_{ab}(\th)}\;,
\sspace S_{ab}(0) =\epsilon_{ab}\;,\;\;\epsilon_{ab}\in \{\pm 1\}\;,
\;\;\epsilon_{aa}=-1\;.
\ee
Given a bootstrap $S$-matrix with these properties the form factors
associated with $S$ are defined to be solutions of a recursive
system of form factor equations\cite{KarWeisz,Smir1}. These
equations admit an algebraic solution scheme in the following sense: There
exists a doublet of form factor algebras $F_{\pm}(S)$ \cite{MN1} such that
`T-invariant' linear functionals $f^{\pm}$ over $F_{\pm}(S)$ define
`pre-'form factors. Their sum $f=f^+ +f^-$ then automatically solves all
the form factors equations, except the bound state residue axiom. Here we
consider a deformation of this construction, motivated by the
considerations outlined in the introduction. We first define an algebra
$F^{(\beta)}(S)$ and then by specialization $F_{\pm}^{(\beta)}(S)$.
In the case of a diagonal $S$-matrix the former is defined as follows:
$F^{(\beta)}(S)$ is an associtive algebra with generators
$t_a^{\pm}(\th),\;\th \in \C,\;W_a(\th),\;0\leq Im\,\th \leq \beta$,
a unit $\1$ and the generators $P_{\mu}\;,\epsilon_{\mu\nu}K$ of the 1+1
dimensional Poincar\'e algebra. Execpt for $P_{\mu}$ all generators
transform as scalars under the action of the Poincar\'e group. The
defining  relations of $F^{(\beta)}(S)$ then are:

\noindent (T) The operators $\,t_1^{\pm}(\th),\ldots,t_n^{\pm}(\th)\,$
generate a direct product of abelian algebras, are $2\pi i$-periodic
and satisfy $t_{\bar{a}}^{\pm}(\th +i\pi)\,t_a^{\pm}(\th)=\1$. The
mixed products $t^+_a(\th_0)t^-_b(\th_1)$ satisfy
$$
S_{ab}(\th_{10}-2\pi i+i\beta)\,t^+_a(\th_0)t^-_b(\th_1)=
S_{ab}(\th_{10})\,t^-_b(\th_1)t^+_a(\th_0)\;.
$$
Further there are linear exchange relations between the
$t^{\pm}$ and the $W$-generators
\vspace{4mm}
\eql{TW}
\bas\jot5mm
&& t_a^+(\th_0)\,W_b(\th_1) = S_{ab}(\th_{01}+2\pi i-i\beta)\,
W_b(\th_1)\,t_a^+(\th_0)\;,\\
&& t_a^-(\th_0)\,W_b(\th_1) = S_{ab}(\th_{01})\,
W_b(\th_1)\,t_a^-(\th_0)\;,
\eas
valid for generic rapidities. The $W$-generators so far are defined
only in the strip $0\leq Im\,\th \leq \beta$. The extension to other
strips $\beta k\leq Im\,\th\leq \beta (k+1),\;k\in \Z$ is done by
repeated use of the relation
\vspace{4mm}
\eql{S}
\bas\jot5mm
&& W_{\bar{a}}(\th)t^+_a(\th +i\beta -i\pi) =
t^-_a(\th +i\beta -i\pi) W_{\bar{a}}(\th+i\beta)\;.
\eas
Finally impose
\vspace{4mm}
\eql{WW}
\bas \jot5mm
W_a(\th_1)\,W_b(\th_2) \is S_{ab}(\th_{12})\;
W_b(\th_2)\,W_a(\th_1)\;,\sspace Re\,\th_{12} \neq 0\;.
\eas
This concludes the definition of the algebra $F^{(\beta)}(S)$. The
product of $W$-generators is defined only when all relative rapidities
have a non-vanishing real part. For relative rapidities that are purely
imaginary, the product of $W$-generators contains simple poles.
The algebra  $F^{(\beta)}(S)$ in which the $W$-generators in addition
satisfy the relation (R$\pm$) below will be denoted by
$F^{(\beta)}_{\pm}(S)$,
respectively. It is convenient to use different symbols $W^+_a(\th)$ and
$W^-_a(\th)$ for $W$-generators satisfying (R$+$) and (R$-$),
respectively. The residue conditions then read
\vspace{4mm}
\eql{R$\pm$}
\bas \jot5mm
&& 2\pi i\,\mbox{res}[W^+_a(\th -i\pi)\,W^+_b(\th)]=
- \delta_{\bar{a}b}\;,\\
&& 2\pi i\,\mbox{res}[W^-_a(\th +i\pi)\,W^-_b(\th)]=
- \delta_{\bar{a}b}\;.
\eas
We shall refer to the algebra $F^{(\beta)}_{\pm}(S)$ as the deformed
{\em form factor doublet}. Implicit in these definitions, of course,
is the presupposition that the above relations define a consistent
algebra. Because of the $\beta$-deformation this does not follow from the
results of \cite{MN1}, but can be proved in a similar way. We proceed with
a number of comments on the structure of  $F^{(\beta)}_{\pm}(S)$. First
note that by means of (S) and (TW), the residue conditions (R$\pm$) imply
\ba
&& 2\pi i\,\mbox{res}[W^+_a(\th +i\beta -i\pi)\,W^+_b(\th)]=
\delta_{\bar{a}b}\, e_a(\th +i\beta -2\pi i)\;,\nonum
&& 2\pi i\,\mbox{res}[W^-_a(\th +i\pi-i\beta )\,W^-_b(\th)]=
-\delta_{\bar{a}b}\,S_{aa}(i\beta)\,e_{\bar{a}}(\th -\pi i)\;.
\ea
where $e_a(\th)=t^-_{\bar{a}}(\th +i\pi)t^+_a(\th)$. For $\beta =2\pi$
this is a central element of $F^{(2\pi)}(S)$ \cite{MN1}; for generic
$\beta$ it still commutes with the $W$-generators. For
generic $\beta$ one could in principle identify $W^+_a(\th)$ and
$W^-_a(\th)$ without running into inconsistencies. It is only in the
limit $\beta\ra 2\pi$ that (R$+$) and (R$-$) would then be in conflict
with the second and first Eqn.~(2.3), respectively.
A second remark concerns the inclusion of bound state poles. Recall that
multiple products of  $W$-generators have been defined only in cases where
at most one relative rapidity is purely imaginary. The extension of the
product to cases where two or more relative rapidities are purely imaginary
is needed for an algebraic implementation of the bound state residue
axiom. We shall not attempt to give a complete discussion here.
To emphasize the point of the condition $ii.$ in the introduction let us
note the residue condition corresponding to simple poles in the
$S$-matrix
\vspace{4mm}
\eql{B1}
\bas \jot5mm
&& 2\pi i\,\mbox{res}[W_a(\th +i\eta(a))\,W_b(\th+i\eta(b))]=
\Gamma_{abc}\,W_{\bar{c}}(\th +i\eta(c)+i\pi)\;,
\eas
where $W$ stands for either $W^+$ or $W^-$ and $\Gamma_{abc}$ is
a constant. Similar conditions would be required for
higher order poles in the $S$-matrix. Their mutual consistency is
non-trivial and will reflect the closure of the $S$-matrix bootstrap.
In addition, the product defined in terms of such residue operations
will in general not be associative, so that the `nesting' of pairs
has to be specified.

Consider now linear functionals over the algebra $F^{(\beta)}(S)$
that are `$T$-invariant' i.e. which satisfy
\be
f^{(\beta)}(X\,t^+_a(\th)) = f^{(\beta)}(X)\;,\sspace
f^{(\beta)}(t_a^-(\th)\,X) = \omega(a)\,f^{(\beta)}(X)\;,
\ee
where $X\in F^{(\beta)}(S)$ has rapidities separated from
$\th$ and $\omega(a)$ is a phase. Functionals $f^{(\beta)}$ corresponding
to non-trivial local operators in addition satisfy $f^{(\beta)}(\1)=0$.
For $\beta =2\pi$ the sum of two such functionals over $F_{\pm}(S)$,
respectively, yields solutions of the form factor equations.
For generic $\beta$ the most important equations for the deformed form
factors are:

\noindent (1) Relation between In and Out states: This is the same
as in the undeformed case. For later use we remark that the action
of the underlying antilinear anti-involution \cite{MN1} is modified
on the $t^{\pm}_a(\th)$ generators. Explicitely
\ba
&& \sigma(W_a(\th))=W_a(\th^*)\;,\sspace \sigma(t^{\pm}_a(\th))=
t^{\mp}_a(\th^*+i\beta -2\pi i)\;.
\ea
\noindent (2) Undeformed exchange relations (1.3).

\noindent (3) $i\beta$-deformed KZE: These are the equations (1.2). If all
relative rapidities have a nonvanishing real part, the $k=n$ equation can
be rewritten in the form
\be
f^{(\beta)}_{a_n\ldots a_1}(\th_n+i\beta,\th_{n-1},\ldots,\th_1) =
f^{(\beta)}_{a_n\ldots a_1}(\th_{n-1},\ldots,\th_1,\th_n)\;.
\ee
Interpreting the rapidities as time variables
this equation has the form of a KMS condition for a thermal n-point
function with inverse temperature $\beta$. Originally it was this
equation, defined for real $\th_n,\ldots, \th_1$ (and $\beta =2\pi$),
that was taken as one of the form factor axioms \cite{Smir1}.

\noindent (4) Deformed kinematical residue equations: The construction in
\cite{MN1} to represent form factors as a sum of two terms $f=f^+ +f^-$
satisfying simple residue equations, remains valid. For
$X^{\pm}= W^{\pm}_{a_n}(\th_n)\ldots  W^{\pm}_{a_1}(\th_1)\in
F_{\pm}^{(\beta)}(S)$ set
$f^{(\beta,\pm)}_{a_n\ldots a_1}(\th_n,\ldots,\th_1):= f^{\pm}(X^{\pm})$.
{}From (R$\pm$) one then finds
\ba
&& 2\pi i\,\mbox{res}\,f^{(\beta,\pm)}_{a_n\ldots a_{k+1}a_k\ldots a_1}
(\th_n,\ldots,\th_k\mp i\pi,\th_k,\ldots,\th_1) \nonum
&&\sspace = -\delta_{\bar{a}_{k+1}a_k}\;
f^{(\beta,\pm)}_{a_n\ldots a_{k+2}a_{k-1}\ldots a_1}
(\th_n,\ldots,\th_{k+2},\th_{k-1},\ldots ,\th_1)\;,
\ea
as in the undeformed case. The reversed residue conditions (2.3)
are $\beta$-dependent and imply
\ba
&& 2\pi i\,\mbox{res}\,
f^{(\beta,+)}_{a_n\ldots a_{k+1}a_k\ldots a_1}
(\th_n,\ldots,\th_k+i\beta -i\pi,\th_k,\ldots,\th_1)
=\delta_{\bar{a}_{k+1}a_k}\bigg[\prod_{j>k+1}S_{a_ka_j}(\th_{kj}+i\beta)
\nonum
&&\sspace \times \prod_{j<k}S_{a_ka_j}(\th_{kj}+2\pi i-i\beta)\bigg]
\;f^{(\beta,+)}_{a_n\ldots a_{k+2}a_{k-1}\ldots a_1}
(\th_n,\ldots,\th_{k+2},\th_{k-1},\ldots ,\th_1)\;,
\ea
together with a similar Eqn.~for $f^-$. By considering the sum
$f=f^+ +f^-$
one can also work out the deformed counterpart of the kinematixal residue
axiom. From (2.7), (2.8) it is clear that for generic $\beta$
the kinematical residue axiom splits up into {\em two} sets of equations
of the form (2.7)(lower case) and (2.8).

\noindent (5) Bound state residue axiom: As indicated before we shall not
attempt here to give a complete discussion of the bound state residue
axiom. The point to emphasize however is that by construction the
conserved charge bootstrap principle will be preserved.

All of the above construction has a direct deformed counterpart also in the
case of a non-diagonal $S$-matrix. A special feature of the diagonal case
is that the KZE admit the following factorized solutions%
\footnote{There are also some non-diagonal theories where this seems to
be the case.}
\be
f^{(\beta,\pm)}_{a_n\ldots a_1}(\th_n,\ldots, \th_1) =
(-)^{n-1}\,K^{(\beta,\pm)}_{a_n\ldots a_1}(\th_n,\ldots,\th_1)\,
\prod_{k>j}\frac{F^{(\beta)}_{a_ka_j}(\th_{kj})}%
{ F^{(\beta)}_{a_ka_j}(\mp i\pi)\,
\sh(\th_{kj}\pm i\pi)\frac{\pi}{\beta} }\;,
\ee
where $F^{(\beta)}_{ab}(\th)$ is the deformed minimal form
factor defined below.
$K^{(\beta,\pm)}_{a_n\ldots a_1}(\th_n,\ldots,\th_1)$
are totally symmetric functions of $\th_n,\ldots,\th_1$ that are
$i\beta$-periodic in each variable and which contain the necessary
bound state poles. Such functions can easily (re-)produced as
correlators of a collection of free fields. For the part carrying
the non-trivial monodromy -- here the product of the minimal
form factors -- this is less obvious. A realization in terms of
generalized vertex operators requires that the quantity in question admits
a factorization into a product of exponentials. Such factorizations are
indeed available for the minimal form factor.
\vspace{-2mm}

\newsection{The minimal form factor as a product of exponentials}
The deformed minimal form factor $F^{(\beta)}_{ab}(\th)$ is uniquely
characterized by the following properties:
\begin{itemize}
\item[(i)]$\Fb(\th)$ is analytic in $0< Im\,\th\leq \beta/2$
and has no zeros in this range. The meromorphic continuation to strips
$S^{(\beta)}_k,\;k\neq 0,\;S^{(\beta)}_k=\{\th\in\C\;|\;k\beta/2 <
Im\,\th \leq (k+1)\beta/2 \}$ is done by means of (ii).
\item[(ii)]
The following equations hold
\be
\Fb(\th)=S_{ab}(\th)\Fb(-\th)\;,\sspace
\Fb(\th+i\beta)=F^{(\beta)}_{ba}(-\th)\;.
\ee
\item[(iii)] Normalization: The limit
$\lim_{|\th|\ra \infty}\Fb(\th)$ exists and $\Fb(i\beta/2)=1$.
\end{itemize}
Following \cite{KarWeisz} we can write down the unique
solution in the following form
\ba
F^{(\beta)}_{ab}(\th) \is \exp\left\{
\frac{\ch^2(\frac{\th\pi}{\beta})}{i\pi}
\int_0^{\infty}dt
\frac{\tanh\frac{t}{2}\;\ln S_{ab}(\frac{\beta}{2\pi}t)}%
{\ch t-\ch\frac{2\pi}{\beta}\th}
\right\}\;,\sspace 0<Im\,\th \leq \beta\;,\nonum
              \is \exp\left\{
              \frac{\ch^2(\frac{\th\pi}{\beta})}{i\pi}
\int_0^{\infty}dt
\frac{\tanh\frac{t}{2}\;\ln S_{ab}(\frac{\beta}{2\pi}t)}%
{\ch t-\ch\frac{2\pi}{\beta}\th}
\right\}\;S_{ab}({\textstyle\frac{2\pi}{\beta}}\th)^{1/2}\;,
\sspace \th\in\R\;.
\ea
The solution simplifies if one assumes that the scattering phase (2.2)
allows for an integral representation of the form
\be
\delta_{ab}(\th)=-\int_0^{\infty}\frac{dt}{t}
h_{ab}(t) \sin\frac{t\th}{\pi}\;,\sspace
0<|Im\,\th|< \sigma_0\leq \pi\;,
\ee
where $\sigma_0$ is the position of the first bound state pole
(i.e~ $S_{ab}(i\sigma)$ is analytic for $0<\sigma <\sigma_0$)
and $h_{ab}$ is subject to the following conditions
\begin{itemize}
\item[--] $h_{ab}$ is real on the real axis and has a
meromorphic continuation off the real axis. All poles
are simple and are given by $\{t=\pm i\pi k,\;k\in E\}$
for some subset $E\subset \N$.
\item[--] $h_{ab}(t) =h_{ba}(t) =h_{ab}(-t)\;,\sspace
|h_{ab}(0)|<\infty.$
\item[--]$ |h_{ab}(t)|\ra e^{-|Re\,t|d_{ab}}\;,\;\;
|Re\,t|\ra \infty$ for constants $\;0<d_{ab}< 1$.
\end{itemize}
In the last condition we assumed that a possible constant term
$k_{ab}\in 2\Z$ in $h_{ab}(t)$ has been split off. Since
$\int_0^{\infty}\frac{dt}{t}\sin\frac{\th t}{\pi}=
\frac{\pi}{2}\mbox{sign}\,\th$ for $\th\neq 0$, this amounts to the
identification $\epsilon_{ab}= e^{\pm i\pi k_{ab}/2}$. The integral
representation (3.3) is
also convenient to derive series expansions of the scattering phase on
and off the imaginary $\theta$-axis. Since the constant term of $h_{ab}$
has been split off the fall-off properties of $h_{ab}$ are such that
series expansions for $\delta_{ab}(\theta)$ are available just by suitable
deformation of the integration contour. This leads to
\be
\delta_{ab}(\th)=\pm \sum_{n\in P}
\frac{d_{ab}(n)}{n}\;e^{\mp n\th}\;,\sspace
\pm Re\,\th>0,\;\;0\leq Im\,\th< \sigma_0\;,
\ee
where $d_{ab}(n)= i\,\mbox{res}_{t=i\pi n}h_{ab}(t)=
-i\frac{h_{ab}(i\pi n)^2}{h'_{ab}(i\pi n)}$. On the
imaginary axis the expansions (3.4) merge to a Fourier series
\be
\delta_{ab}(i\sigma)=i\sum_{n\in P}
\frac{d_{ab}(n)}{n}\;\sin\,n\sigma\;,\sspace
0<\sigma<2\pi\;.
\ee
So far the $S$-matrix bootstrap equations (2.1) have not been
used. Imposing them on the level of the series expansions
(3.4), (3.5) puts constraints on the expansion coefficients
$d_{ab}(n)$. The most prominent (possibly all) solutions of
these constraints are those descending from Lie algebraic data.
In that case the particles $a=1,\ldots,r$ are associated with
the Dynkin diagram of a simple Lie algebra $g$ and the possible
fusing angles are selected by the condition
\be
\sum_{l=a,b,c}e^{\pm is\eta(l)}\,q_l^{(s)}=0\;,
\ee
where $(q^{(s)}_1,\ldots,q^{(s)}_r)^T$ is the normalized
real eigenvector of the Cartan matrix with eigenvalue
$2(1-\cos\frac{s\pi}{h})$ ($h$: Coxeter number, $s$: exponent).
The coefficients $d_{ab}(n)$ then take the form
\be
d_{ab}(n) =d_n\,q^{(n)}_a q^{(n)}_b\;,
\ee
for real constants $d_n$. The second and third eqn.~in (2.1) are satisfied
by means of $q^{(n)}_{\bar{a}}=(-)^{n+1}q^{(n)}_a$ and (3.6), respectively.
Notice that the coefficients (3.7) vanish unless $n$ is an
exponent of $g$ modulo $h$, so that the summations over $n\in E$ here
correspond to summmations over the set of affine exponents. It has been
shown in \cite{MN1,MN2} that
\be
I^{(n)}_a =c^{n/2}\,\sqrt{\frac{d_n}{n}}\,q_a^{(n)}\;,\;\;\; n\in E
\ee
is the exact eigenvalue of the $n$-th local conserved charge on a single
particle state of type $a$ at rapidity zero.

Now return to the deformed minimal form factor (3.2). Expressed in terms
of the function $h_{ab}(t)$ it reads
\be
F^{(\beta)}_{ab}(\th)
 = \left(-i\sh\frac{\pi\th}{\beta}\right)^{k_{ab}/2}\,
\exp\left\{\int_{0}^{\infty}\frac{dt}{t}\,
\frac{ h_{ab}(\frac{2\pi}{\beta}t) }{\sh t}\,
\sin^2(i\pi-\frac{2\pi}{\beta}\th)\frac{t}{2\pi}\right\}\;,
\ee
for $0< Im\,\th <\beta$.
It is sometimes convenient to separate the phase and the
modulus of $\Fb(\th)$. Writing $\th =\vartheta +i\sigma$ one has
in particular: $\arg \Fb(\vartheta,0)=-\frac{\pi}{4}k_{ab}+
\frac{1}{2}\delta_{ab}(\vartheta)$, while for purely imaginary
$\th=i\sigma$ the minimal form factor is real: $\arg \Fb(0,\sigma)=0$.
For the rest of this paper we will {\em formally set $k_{ab}=0$} in order
to simplify the expressions. The dependence on $k_{ab}$ can easily be
restored through eqn.s (2.2), (3.9).

In the following we will derive two types of series expansions for
$\ln F^{(\beta)}_{ab}(\th)$. The first type is valid for all real
positive $\beta$, including $\beta =2\pi$; the second type is valid only
for $\beta/2\pi$ irrational. On the imaginary $\th$-axis it is natural to
consider Fourier series, while off the imaginary axis
expansions in $e^{\pm \th}$ for $\pm Re\,\th<0$ will be
used. Let us first consider expansions valid for all $\beta>0$.
We claim that
\be
\ln \Fb(i\sigma) = \sum_{n\geq 1}
[\cos{\textstyle \frac{2\pi n}{\beta}\sigma }+(-)^{n+1}]\;
\Fb(n)\;,\sspace 0<\sigma <\beta\;,
\ee
where
\be
F^{(\beta)}_{ab}(n)=\int_0^{\infty}ds\,
e^{-ns}\delta_{ab}({\textstyle \frac{\beta}{2\pi}}s)\;,
\ee
or in the representation (3.3)
\be
\Fb(n) =-\int_0^{\infty}dt
\frac{h_{ab}(\frac{2\pi t}{\beta})}{(\pi n)^2 +t^2}\;.
\ee
Off the imaginary $\th$-axis one has
\be
\ln F^{(\beta)}_{ab}(\th)=\sum_{n\geq 1}
[e^{\pm n\frac{2\pi}{\beta}\th}+(-)^{n+1}]\;
F^{(\beta)}_{ab}(n)\;,\sspace
0<Im\,\th<\beta,\;\;\pm Re\,\th<0\;,
\ee
with the same coefficients (3.11), (3.12). In order to verify Eqn.~(3.10)
first note that
$$
\sum_{n\geq 1}e^{-nt}[\ch n\th +(-)^{n+1}]=
\frac{\tanh\frac{t}{2}\;\ch^2\frac{\th}{2}}{\ch t-\ch\th}\;,
$$
for $|Re\,\th|<Re\,t$. Inserting into Eqn.~(3.2) one can
exchange the order of summation and integration, which yields (3.10)
with coefficients (3.11). In the representation (3.3) a simple
integration yields the coefficients in the form (3.12). The
derivation of (3.13) runs similarly: Combining
$$
\frac{1+e^{-\th}}{(1+e^{-t})(e^t-e^{-\th})}+
\frac{1+e^{\th}}{(1+e^{-t})(e^t-e^{\th})}=
\frac{\tanh\frac{t}{2}\;\ch^2\frac{\th}{2}}{\ch t-\ch\th}
$$
and
$$
\sum_{n\geq 1}[e^{\pm n\th}+(-)^{n+1}]=
\frac{1+e^{\pm\th}}{(1+e^{-t})(e^t-e^{\pm\th})}
$$
one arrives at (3.12).

The expansions (3.10), (3.13) are valid for all $\beta >0$, in
particular for the physical value $\beta =2\pi$.
For practical purposes, however, the expansions (3.10), (3.13) for
$\beta =2\pi$ are not particularly useful. Except for a few simple cases
the expansion coefficients
$F^{(2\pi)}_{ab}(n)$ cannot be evaluated explicitely (or can only
be written in the form of a trigonometric series or trigonometric
polynomials with $n$-dependent order). Moreover one would like to relate
the coefficients $F^{(\beta)}_{ab}(n)$ of $\ln \Fb(\theta)$ and
$d_{ab}(n)$ of $\delta_{ab}(\theta)$. This can be achieved by
taking $\beta$ off $2\pi$. The main observation is that for
$\beta/2\pi$ irrational the double poles of the integrand in (3.9) are split:
By assumption $t\ra h_{ab}(2\pi t/\beta)$ has simple poles at $\{t=\pm
i\beta k/2,\; k\in E\}$ while $t\ra 1/\sh t$ has simple poles
at $\{t=\pm i\pi n,\;n\in \N\}$. Thus, for $\beta/2\pi$ irrational
both sets of poles do not intersect. This means that
all the poles of the integrand are now simple, so that one can
derive series expansions of the desired form just by deforming the
integration contour. On the imaginary $\th$-axis one finds
\ba
\ln\,\Fb(i\sigma) \is \frac{1}{2}\sum_{n\geq 1}\frac{1}{n}
   h_{ab}\left({\textstyle \frac{2\pi i}{\beta}\pi n}\right)
   \left( (-)^n - \cos
   {\textstyle \frac{2\pi n}{\beta}\sigma }\right)\nonum
&\!-\!&\frac{1}{2}\sum_{n\geq 1}\frac{1}{n}
  \frac{d_{ab}(n)}{\sin\frac{\beta}{2}n}\;
  \left(1-\cos {\textstyle n(\frac{\beta}{2}-\sigma)}\right)\;,
\sspace 0<\sigma< \beta\;.
\ea
Off the imaginary axis the result is
\ba
\ln\,\Fb(\th) \is \frac{1}{2}\sum_{n\geq 1}\frac{1}{n}
   h_{ab}\left({\textstyle \frac{2\pi i}{\beta}\pi n}\right)
   \left( (-)^n - e^{\mp 2\pi n\th}\right)\nonum
&\!-\!&\frac{1}{2}\sum_{n\geq 1}
  \frac{d_{ab}(n)}{n\,\sin\frac{\beta}{2}n}\;
  \left(1-e^{\pm n(i\frac{\beta}{2}-\th)}\right)\;,
\nonum && \mbox{for}\;\;\pm Re\,\th>0\;\;\mbox{and}\;\;
0<Im\,\th <\beta\;.
\ea
The advantage of the expansions (3.14), (3.15) is that all the expansion
coefficients are known explicitely in terms of the function $h_{ab}(t)$.
Moreover, the apparently unattractive feature of having a doubling of
terms (with expansion coefficients $h_{ab}(i2\pi^2 n/\beta)$ and
$d_{ab}(n)$, respectively) corresponds to a decomposition
into a part with trivial and with non-trivial monodromy. Before turning
to these issues let us discuss the relation to, and consistency with, the
series expansions (3.10), (3.13) valid for all $\beta >0$. To do so,
consider the expression (3.12) for the coefficients $F^{(\beta)}_{ab}(n)$.
For $\beta/2\pi$ irrational the integrand in (3.12) has again only simple
poles and the integral again can be done by deforming the contour.
Using different choices of the contour one obtains the
equivalent expressions
\ba
\Fb(n) \is -\frac{1}{2n}h_{ab}\left(
{\textstyle\frac{2\pi i}{\beta}n\pi}\right)
-\frac{\beta}{2\pi^2}\sum_{k\in E}
\frac{d_{ab}(k)}{n^2 -(k\beta/2\pi)^2}\nonum
\is -\frac{1}{n}h_{ab}\left(
{\textstyle\frac{2\pi i}{\beta}n\pi}\right)
-\frac{\beta}{2\pi^2}\sum_{k\in E}
\frac{d_{ab}(k)}{n(n-k\beta/2\pi)}\nonum
\is -\frac{\beta}{2\pi^2}\sum_{k\in E}
\frac{d_{ab}(k)}{n(n+k\beta/2\pi)}\;,
\ea
for $0<\beta <2\pi$. In particular (3.16) displays the relation between the
coefficients $\Fb(n)$ and $d_{ab}(n)$ searched for.
Notice that only in the last of the expressions (3.16) can one take the
limit $\beta\ra 2\pi$. Inserting the first expression for $\Fb(n)$
into (3.10) one can justify the exchange in the order of
summations. Using then
\bas
&& \sum_{n\geq 1}[\cos{\textstyle \frac{2\pi n}{\beta}\sigma}
                  + (-)^{n+1}]\;
\frac{1}{n^2 - (k\beta/2\pi)^2} \nonum
&&\bspace =\frac{2\pi^2}{\beta}\frac{1}{2k\sin\frac{\beta}{2}k}
\left(1-\cos {\textstyle k(\frac{\beta}{2}-\sigma)}\right)\;,
\sspace 0<\sigma<\beta <2\pi\;,
\eas
for the resummation one recovers (3.14).

Set now
\be
\phi^{(\beta)}_{ab}(\theta) := \sum_{n\geq 1}
  \frac{d_{ab}(n)}{2n\,\sin\frac{\beta}{2}n}\;
  e^{\pm n(i\frac{\beta}{2}-\th)}\;,\;\; \mbox{for}\;\;
\pm Re\,\th>0\;\;\mbox{and}\;\;
0<Im\,\th <\beta\;,
\ee
and let $\nu^{(\beta)}_{ab}(\th)$ denote the rest of the terms in (3.15),
so that $\ln \Fb(\th) =\nu^{(\beta)}_{ab}(\th)+
\phi^{(\beta)}_{ab}(\th)$. One then verifies that
\ba
\nu^{(\beta)}_{ab}(\th+i\beta) = \nu^{(\beta)}_{ab}(\th)\;,\sspace &&
\nu^{(\beta)}_{ab}(\th) =\nu^{(\beta)}_{ab}(-\th)\;,\nonum
\phi^{(\beta)}_{ab}(\th+i\beta) =\phi^{(\beta)}_{ab}(-\th)\;,\sspace &&
\phi^{(\beta)}_{ab}(\th) =\phi^{(\beta)}_{ab}(-\th)+i\delta(\th)\;,
\;\; Re\,\th \neq 0\;,
\ea
using the expansion (3.4) for $\delta(\theta)$.
This means that $\phi^{(\beta)}_{ab}(\th)$ carries all the non-trivial
monodromy and $\exp \phi^{(\beta)}_{ab}(\th)$ is itself a solution to the
conditions $(i)$ and $(ii)$ on the $\beta$-deformed minimal
form factor. The only feature it does not have is a well-defined
limit as $\beta\ra 2\pi$. The first term $\exp \nu^{(\beta)}_{ab}(\th)$
has trivial monodromy, has again no $\beta \ra 2\pi$ limit, but is
designed s.t. the product of both factors does have a limit, which by
construction coincides with the original minimal form factor.

Eqn.s (3.9),(3.14),(3.15) also exhibit the difference to the regularization
of the form factors used in \cite{Kyoto,Luk}. From the viewpoint of the
continuum theory, the construction in \cite{Kyoto} approximates
integrals of the form (3.9) (for $\beta =2\pi$) by a Riemann sum, which is
natural in the context of lattice models. The $S$-matrix of the asymptotic
particles in the scaling limit is only indirectly related to the
$R$-matrix, so that the analogue of the condition $ii.$ in the
introduction does not exist (or, if $S$ is replaced with $R$, is empty).
In the context of the form factor bootstrap, however, a regularization that
violates the condition $ii.$ in the introduction is problematic. A
deformation  of the $S$-matrix will in general also affect the structure of
its bound state poles, in which case the consistency (3.6) with the spins
of the local conserved charges (i.e. Zamolodchikov's `conserved charge
bootstrap' principle) is spoiled.


\newsection{Free field realization of \mbox{\boldmath$F^{(\beta)}(S)$}}
The feature (3.18) of the series expansion (3.15) suggests to
rewrite equation (2.9) in the following form
\be
f^{(\beta,\pm)}_{a_n\ldots a_1}(\th_n,\ldots, \th_1) =
\widetilde{K}^{(\beta,\pm)}_{a_n\ldots a_1}(\th_n,\ldots,\th_1)\,
\prod_{k>j}e^{\textstyle \phi^{(\beta)}_{a_ka_j}(\th_{kj})}\;,
\ee
where
$$
\widetilde{K}^{(\beta,\pm)}_{a_n\ldots a_1}(\th_n,\ldots,\th_1)=
(-)^{n-1}K^{(\beta,\pm)}_{a_n\ldots a_1}(\th_n,\ldots,\th_1)\,
\prod_{k>j}\frac{e^{\textstyle \nu^{(\beta)}_{a_ka_j}(\th_{kj})}}%
{F^{(\beta)}_{a_ka_j}(\mp i\pi)\,\sh(\th_{kj}\pm i\pi)\frac{\pi}{\beta}}
$$
can be realized as a correlator
\be
\widetilde{K}^{(\beta,\pm)}_{a_n\ldots a_1}(\th_n,\ldots,\th_1)=
\bra \cO| Q^{\pm}_{a_n}(\th_n)\ldots Q^{\pm}_{a_1}(\th_1)|0\ket \;,
\ee
of $i\beta$-periodic fields $Q^{\pm}_a(\th)$ with trivial monodromy.
The part carrying the non-trivial monodromy will be realized as
\be
\bra \rho_I(W_{a_n}(\th_n))\ldots  \rho_I(W_{a_1}(\th_1))\ket =
\prod_{k>j}e^{\textstyle \phi^{(\beta)}_{a_ka_j}(\th_{kj})}\;,
\ee
where the operators $\rho_I(W_a(\th))$ and $\rho_I(t^{\pm}_a(\th))$
form a realization of $F^{(\beta)}(S)$. In total, the realization of
the deformed form factor algebra $F_{\pm}^{(\beta)}(S)$ is a tensorproduct
\be
\rho(W^{\pm}_a(\th))= Q^{\pm}_a(\th)\otimes \rho_I(W_a(\th))\;,\;\;\;
\rho(t^{\pm}_a(\th)) =\1\otimes \rho_I(t^{\pm}_a(\th))\;,
\ee
where both factors act on different free field Fock spaces.
It remain to construct the realization $\rho_I$ of  $F^{(\beta)}(S)$.
It turns out that the latter can be constructed
entirely in terms of the local conserved charges $I^{(n)},\;n\in E$
together with their images under the antilinear anti-involution
$\sigma$ defined in (2.5). The starting point is the relation \cite{MN1}%
\footnote{For notational simplicity we denote $\rho_I(t^{\pm}_a(\th))$
by $t^{\pm}_a(\th)$ etc. in the following. Since we redefined $S$ by
assuming $k_{ab} =0$ zero modes are absent.}
\be
t^+_a(\th) =\exp\left\{\pm i\sum_{n\in E}\frac{e^{\mp\th n}}{c^n}
\,I^{(n)}_a I^{(n)}\right\}\;,
\ee
valid for $\pm Re(\th -\th_i)>0\;,\sigma_0>Im\,\th \geq 0$, acting on a
multiparticle state with real rapidities $\th_1,\ldots, \th_n$. The operator
$t_a^-(\th)$ can then be computed from the involution (2.5)
\be
t^-_a(\th) =\exp\left\{\pm i\sum_{n\in E}
\frac{e^{\pm(\th+i\beta)n}}{|c|^n}\,I^{(n)}_a \sigma(I^{(n)})\right\}\;,
\ee
taking into account (3.8) and the correct matching of branches.
The relations (T) then fix the commutator between
$I^{(n)}$ and $\sigma(I^{(m)})$. One finds $[\dd{x_n},x_m]=\delta_{mn}$
with the definitions
\be
I^{(n)}=\dd{x_n}\;,\sspace \sigma(I^{(n)})=
\mp i|c|^n(1-e^{\mp i\beta n})\,x_n\;,
\ee
where the sign options refer to that in (4.5),(4.6). The
relations (TW) are equivalent to
\bas
[\ln t^+_a(\th_0)\,,\,W_a(\th_1)] \is i\delta_{ab}(\th_{01}-i\beta)\,
W_a(\th_1) \;,\nonum
[\ln t^-_a(\th_0)\,,\,W_a(\th_1)] \is - i\delta_{ab}(\th_{10})\,
W_a(\th_1) \;,
\eas
so that
\bas
[\dd{x_n}\,,\,W_a(\th)] \is I^{(n)}_a\,e^{\mp(\th +i\beta)n}
W_a(\th_1) \;,\nonum
[x_n\,,\,W_a(\th)] \is \frac{\pm i}{c^n}
\frac{e^{\mp n\th}}{1-e^{\pm i\beta n}}\,I^{(n)}_a\,W_a(\th) \;.
\eas
This implies
\be
W_a(\th) =\exp\left\{\sum_{n\in E}I^{(n)}_a
e^{\pm n(\th +i\beta)}\,x_n\right\}
\exp\left\{\mp i\sum_{n\in E}\frac{I^{(n)}_a}{c^n}
\frac{e^{\mp n\th}}{1-e^{\pm i\beta n}}\,\dd{x_n}\right\}\;.
\ee
One can now check that the operators (4.5), (4.6) and (4.8) also satisfy
the relations (S) and (WW) so that they yield a realization of
$F^{(\beta)}(S)$. In particular one finds
\be
W_a(\th_1)W_b(\th_2)= :W_a(\th_1)W_b(\th_2):\,
e^{\textstyle \phi^{(\beta)}_{ab}(\th_{12})}\;,\;\;\;Re\,\th_{12}\neq 0\;.
\ee
where $:\;\;:$ indicates normal ordering in the Heisenberg algebra
generated by $\dd{x_n},\;x_n$ and $\phi^{(\beta)}_{ab}(\th)$ is defined
in (3.17). Thus, if $\bra\;,\;\ket:=\bra \;,\;\ket_{\omega}$ denotes the
canonical sesquilinear form (contravariant w.r.t $\omega(x_n)=\dd{x_n}$)
on the Fock space $\C[x_1,x_2,\ldots]$, equation (4.3) follows, which
completes the construction.

We add two comments:
In view of equation (2.6) one might expect the appearence of
trace functionals as in \cite{Kyoto,Luk}. Because of the Clavelli-Shapiro
formula \cite[App.C1]{CS} however the distinction between trace- and vector
functionals is not an intrinsic one. In upshot, the result
\cite[App.C1]{CS} can be rephrased as follows: If in a Heisenberg algebra
$[d\,,\,d^{\dagger}]=1$ one uses the non-standard antilinear
anti-involution $\sigma(d) = \frac{q}{1-q} d^{\dagger}$ to define a
sesquilinear form contravariant w.r.t.~it (so that e.g.
$\bra d,d\ket_{\sigma} =q/(1-q)$) the resulting expectation values can
be interpreted as trace functionals w.r.t.~$q^{d^{\dagger}d}$ and the
sesquilinear form  contravariant w.r.t $\omega(q) =q^{\dagger}$. More
generally, this phenomenon is related to the GNS construction. A second
comment concerns the possible physical significance of the QFTs defined
in terms of the deformed form factors. Since the scattering theory - and
hence the bound state structure - is preserved, the deformation affects
only the kinematical arena. One might therefore expect that the
$\beta$-deformed QFTs admit an interpretation as QFTs living on some
deformed spacetime.


\end{document}